\documentclass[twocolumn,showpacs,preprintnumbers,amsmath,amssymb,superscriptaddress,pra]{revtex4}
\usepackage{dcolumn}
\usepackage{bm}
\usepackage{graphicx}
\newcommand{\ket}[1]{\displaystyle{|#1\rangle}}

\begin{document}

\title{Interferometry using spinor Bose-Einstein condensates}

\author{R. Vasile}
\affiliation{Department of Physics and Astronomy, University of
Turku, 20014 Turun Yliopisto, Finland}
\author{H. M\"akel\"a}
\affiliation{Department of Physics and Astronomy, University of
Turku, 20014 Turun Yliopisto, Finland}
\affiliation{Department of Physics, Sofia University, James Bourchier 5 blvd, 1164 Sofia, Bulgaria}
\author{K.-A. Suominen}
\affiliation{Department of Physics and Astronomy, University of
Turku, 20014 Turun Yliopisto, Finland}

\date{\today}

\begin{abstract}
We study the time-evolution of an optically trapped spinor
Bose-Einstein condensate under the influence of a dominating
magnetic bias field in the $z$-direction, and a perpendicular
smaller field that couples the spinor states. We show that if the
bias field depends quadratically on time, the relative phases of the
spinor components affect the populations of the final state. This
allows one to measure the differences in the time-evolution of the
relative phases, thereby realizing a multi-arm interferometer in a
spinor Bose-Einstein condensate.
\end{abstract}


\maketitle

\section{Introduction}\label{Sec:Intro}

The realization of Bose-Einstein condensation (BEC) in optically
trapped ultracold atomic gases~\cite{Opttrap} has made it possible
to study the spinor dynamics of multicomponent, or spinor, BECs. The
theoretical framework describing their dynamics provides a set of
coupled time-dependent Gross-Pitaevskii equations for the different
components of the order parameter~\cite{Machida,Ho,Bigelow}. The
internal dynamics can be studied experimentally by taking advantage
of the coupling between the hyperfine spin of the condensed atoms
and an external magnetic field. By choosing appropriately the
magnetic field it is possible to control and detect the state of the
spinor condensate. The time scale characterizing changes in the
external magnetic field can be made much shorter than the time scale
of the spin mixing dynamics~\cite{Bigelow2}. Then the latter can be
neglected, resulting in a simplified theoretical description.

One possible experimental configuration consists of an initially
strong magnetic field in the $z$-direction (bias field) and a weaker
perpendicular field (coupling field) in the $xy$-plane. The spinor
states become Zeeman-shifted by the bias field, and adjacent states
are coupled resonantly when the bias field actually crosses a
zero-value point (i.e., reverses its direction). Then the spinor
dynamics can be described by the Landau-Zener level crossing model
(LZ)~\cite{Landau32,Zener32,Majorana32}. Note that this is not the
only possible scenario, and towards the end of the paper we discuss
other possibilities.

In this paper we consider a level crossing model known as the
parabolic model, in which the energy levels have a quadratic
time-dependence. Depending on the value of the parameters the energy
levels either do not cross, cross at one point, or cross twice. In
the double crossing case the propagator can be, to a very good
approximation, obtained by applying the LZ model once at each
crossing and taking into account the dynamical phase accumulated by
the different spinor components between the
crossings~\cite{Suominen92}. The standard version of the parabolic
model involves two energy levels. However, we are interested in
applying it in the context of spinor Bose-Einstein condensates,
which have an odd number of energy levels. We thus generalize the
parabolic model to these situations using the Majorana
representation~\cite{Majorana32,Bloch45,Vitanov97,Xia08}.

The new feature we want to point out is that by using the parabolic
model we can realize a multi-arm interferometer in a spinor
Bose-Einstein condensate. The role of the two crossings is to mix
spinor components, thus they work as beam splitters in quantum
optics~\cite{GerryKnight}. Between the crossings the spinor state
changes due to the external magnetic field and the time-evolution
determined by the Gross-Pitaevskii equations. The original parabolic
model does not take into account interparticle interactions or other
additional phase effects, so in this paper we analyze their role.

The paper is organized as follows. In Section~\ref{Sec:MajoLZ} we
introduce the Majorana representation and apply it to the
Landau-Zener model. In Section~\ref{Sec:Para} we describe the
two-level parabolic model and its generalization to higher
dimensions. Finally in Section~\ref{Sec:Interf} we show how the
parabolic model can be used to detect the phase differences of the
spinor components developed between the crossings.

\section{The Majorana representation and the Landau-Zener model}\label{Sec:MajoLZ}

Any two-level system Hamiltonian, with explicit time dependence, can
be written as
\begin{equation}\label{BBB}
   H^{(2)}(t)=A(t)I^{(2)}+\sum_{i=1}^{3}D_{i}(t)S^{(2)}_{i},
\end{equation}
where $I^{(2)}$ is the identity matrix and $S^{(2)}_{i}=\sigma_i/2$
is the spin-$1/2$ operator in the $i$th direction, given in terms of
the Pauli matrices $\sigma_i$. The time evolution determined by this
Hamiltonian is obtained by solving the time-dependent Schr\"odinger
equation
\begin{equation}
   i\hbar\frac{\partial\psi(t)}{\partial t}=H^{(2)}(t)\psi(t).
\end{equation}
The solution can be written in terms of an evolution operator, or
propagator $U^{(2)}(t,t_0)$, which generates the state at time $t$
when the state at a previous time $t_0$ is given. The general
unitary form of the propagator for a two-level system is
\begin{equation}
\label{U2}
   U^{(2)}(t,t_0)=\left(
                 \begin{array}{cc}
                   \alpha & \beta \\
                   -\beta^{*} & \alpha^{*} \\
                 \end{array}
               \right),\quad |\alpha|^2+|\beta|^2=1.
\end{equation}
Its exact or approximate expression is known only for a few specific
cases of Hamiltonian~\eqref{BBB}. When these models are generalized
into a higher number of levels forming the spinor components, the
elements of the corresponding propagator can be given in terms of
the elements of $U^{(2)}$ using the Majorana representation.
Explicitly, if the Hamiltonian of an $n$-level spinor system is
\begin{equation}\label{CCC}
   H^{(n)}(t)=A(t)I^{(n)}+\sum_{i=1}^{3}D_{i}(t)S^{(n)}_{i},
\end{equation}
with $A$ and $D_i$ as in Eq.~\eqref{BBB}, and $S^{(n)}_{i}$ are the
spin $S$ ($S=2n+1$) angular momentum operators, then the elements of
the propagator $U^{(n)}(t,t_0)$ are given by
\begin{eqnarray}\label{Majo}
   \label{Ugeneral} \nonumber
   &&(U^{(n)})_{ij}=\sum_{k=0}^{n-i}\sum_{l=1}^{i-1}\alpha^{k}\beta^{n-i-k}
   (-\beta^{*})^{l}   (\alpha^{*})^{i-l-1}\\
   &&\times\frac{\sqrt{(n-i)!(i-1)!(k+l)!(n-k-2)!}}{k!l!(n-i-k)!(i-l-1)!},
\end{eqnarray}
where we have not explicitly denoted the time-dependence. This
result has been derived in~\cite{Majorana32,Bloch45} (see
also~\cite{Vitanov97,Meckler58}). We emphasize that
Eq.~\eqref{Ugeneral} is valid only if the Hamiltonian of an
$n$-level system can be written as in Eq.~\eqref{CCC}. This is true
for bosonic atoms in an external magnetic field as long as we have
only the linear Zeeman shift.

Let us now apply the Majorana representation to the Landau-Zener
model~\cite{Landau32,Zener32,Majorana32}, i.e., the linear crossing
model. The LZ Hamiltonian is
\begin{equation}\label{LZH}
   H_{LZ}^{(2)}(t)=\left(
         \begin{array}{cc}
           \lambda t & V_0 \\
           V_0 & -\lambda t\\
         \end{array}
       \right),
\end{equation}
where $\lambda$ and $V_0$ are positive constants. Many different
approaches have been developed to study the dynamics of such a
two-level system. In this paper we work in the instantaneous
eigenstate basis, i.e.~the adiabatic basis of
Hamiltonian~\eqref{LZH}, since it allows us to use the results of
Ref.~\cite{Suominen92} directly. In the LZ model as well as in the
parabolic double crossing model the state vectors of the adiabatic
basis coincide (apart from possible sign differences) with those of
the original diabatic basis of Eq.~\eqref{LZH} far from the crossing
region. If we choose $t_0=-\infty$ and $t=+\infty$, the propagator
in the adiabatic basis is~\cite{Kazantsev1990}
\begin{equation}\label{LZProp}
   U_{LZ}^{(2)}(+\infty,-\infty)=\left(
                     \begin{array}{cc}
                       \sqrt{1-R^2}e^{-i\phi} & -R \\
                       R & \sqrt{1-R^2}e^{i\phi} \\
                     \end{array}
                   \right),
\end{equation}
where
\begin{eqnarray}
   \label{R}
   R&=&\exp\left[-\frac{\pi\Lambda}{2}\right],\\
   \label{Lambda}
   \Lambda &=& \frac{V_0^2}{\hbar\lambda},
\end{eqnarray}
and $\Lambda$ is the Landau-Zener parameter. The Landau-Zener phase $\phi$ appearing in Eq.~(\ref{LZProp}) is
\begin{equation}\label{LZphase}
   \phi=\frac{\pi}{4}+\frac{\Lambda}{2}\ln\left(\frac{\Lambda}{2e}\right)
   +\arg\left[\Gamma(1-\frac{i\Lambda}{2})\right],
\end{equation}
where $\Gamma$ denotes the Euler Gamma function. The value of $\phi$
decreases monotonically from $\pi/4$ to 0 as $\Lambda$ goes from 0
to infinity. Note that the propagator~\eqref{LZProp} is given in the
interaction picture (e.g., Eq.~(5) in Ref.~\cite{Suominen92}). In
the Landau-Zener model (in both diabatic and adiabatic basis) the
energy level separations are initially and finally infinite, which
means that it is convenient to use the interaction picture.

Applying Eq.~\eqref{Majo} we can now evaluate the propagator for any
number $n$ of levels. In the simplest case of $n=3$ we have
\begin{equation}\begin{split}
\footnotesize U^{(3)}_{LZ}(+\infty,-\infty)=\qquad\qquad\qquad\qquad\qquad\qquad\qquad\qquad\qquad\qquad&\\
 \footnotesize\left(
          \begin{array}{ccc}
            (1-R^2)e^{-2i\phi} & -R\sqrt{2(1-R^2)}e^{-i\phi} & R^2 \\
             R\sqrt{2(1-R^2)}e^{-i\phi} & 1-2R^2 & -R\sqrt{2(1-R^2)}e^{i\phi} \\
            R^2  & R\sqrt{2(1-R^2)}e^{i\phi} & (1-R^2)e^{2i\phi} \\
          \end{array}
        \right)&.
\end{split}
\end{equation}

In the following section we consider the special case of the
parabolic model in which two consecutive crossings appear. We work
under the independent crossing approximation (ICA), which states
that the two crossings can be treated as separate linear crossings.
The relevant time scale in a single linear crossing is given by the
Zener time $t_Z$,
\begin{equation}
   t_Z=\left\{
   \begin{array}{ll} \frac{V_0}{\lambda},&\Lambda\gg 1,\\
   \\
   \sqrt{\frac{\hbar}{\lambda}}, & \Lambda\ll 1,
   \end{array}\right.
\end{equation}
where the upper line corresponds to the adiabatic limit and the
lower one to the sudden limit. Then the condition for ICA reads
simply as
\begin{equation}\label{Separation}
   t_{Z}\ll t_c,
\end{equation}
where $t_c$ is the time interval between the
crossings~\cite{Mullen89}. The parameters of the parabolic model
will be chosen to satisfy this requirement (see Fig.~\ref{fig:1}).

\section{Parabolic model}\label{Sec:Para}

\begin{figure}[b]
\begin{center}
\includegraphics[width=8.6cm]{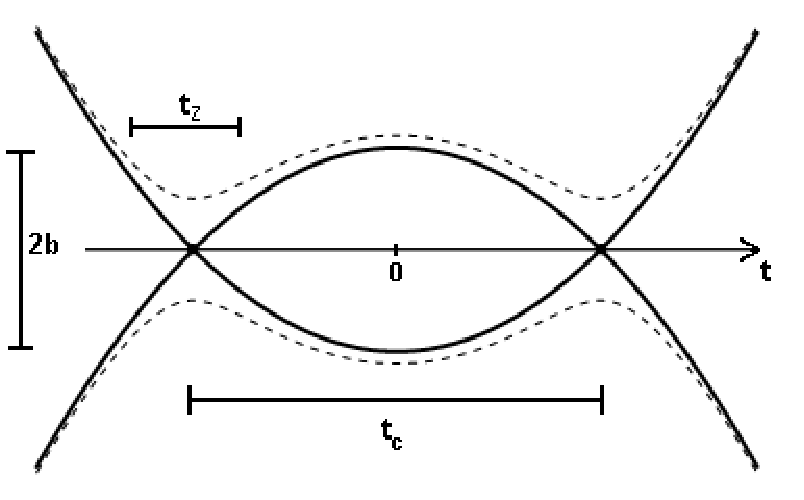}
\end{center}
\caption{The two-level parabolic model. Solid lines are the diabatic
energy levels as a function of time, and dashed lines are the
adiabatic energy levels (featuring an avoided crossings). We choose
the scaled coupling $b>0$ in order to have a double crossing case.
The Zener time $t_Z$ and the crossing separation $t_c$ are also
shown.}\label{fig:1}
\end{figure}

We now discuss the two-level parabolic
model~\cite{Bykhovskii65,Crothers77,Shimshoni91,Suominen92} and its
extensions to a higher number of levels. The Hamiltonian of the
parabolic model is
\begin{equation}\label{ParaH}
   H_{Para}^{(2)}=\left(
         \begin{array}{cc}
           a t^2-b & v \\
           v & -at^2+b\\
         \end{array}
       \right),
\end{equation}
where $a$ and $v$ are positive and $b$ is a real number. The sign of
$b$ determines whether we have a double crossing ($b>0$), a single
crossing ($b=0$), or a tunneling case ($b<0$). In this paper we will
consider only the $b>0$ case (Fig.~\ref{fig:1}). In terms of the
spin-1/2 operators the parabolic Hamiltonian can be written as
$H_{Para}^{(2)}=2v \,S_x^{(2)}+2(at^2-b)S_z^{(2)}$. It is convenient
to scale the system by defining a dimensionless time $\tau=vt/\hbar$
which leads to the new Hamiltonian
\begin{equation}\label{ParaHScaled}
\bar{H}_{Para}^{(2)}=\left(
         \begin{array}{cc}
           \epsilon \tau^2-\mu & 1 \\
           1 & -\epsilon \tau^2+\mu\\
         \end{array}
       \right),
\end{equation}
where $\epsilon=\hbar^2 a/v^3$ and $\mu=b/v$ are dimensionless parameters~\cite{Suominen92}.

Analytic solutions for the propagator
$U^{(2)}_{Para}(+\infty,-\infty)$ have been presented in the
literature~\cite{Shimshoni91,Suominen92} in some limiting cases.
Reference~\cite{Suominen92} provides an analytic solution in the
adiabatic limit, using the interaction picture and the adiabatic
basis representation. If generalized, this solution suggests that
more generally, if condition~\eqref{Separation} is satisfied, so
that the two resonances can be treated as independent linear
crossings (ICA), their contribution can be described by the
Landau-Zener propagator of Eq.~\eqref{LZProp}. Linearizing
$\bar{H}_{Para}^{(2)}$ at the crossings we find that the associated
Zener time is given in terms of the scaled time as
\begin{equation}
   \tau_Z =\left\{
   \begin{array}{ll}
    \frac{1}{2\sqrt{\mu\epsilon}},& 2\sqrt{\mu\epsilon}\ll 1,\\
    \\
    \sqrt{\frac{1}{2\sqrt{\mu\epsilon}}},& 2\sqrt{\mu\epsilon}\gg 1,
    \end{array}\right.
\end{equation}
with the upper line corresponding to the adiabatic and the lower
line to the sudden limit. The distance between the crossings is
$\tau_c=2\sqrt{\mu/\epsilon}$. Thus the ICA
condition~\eqref{Separation} is satisfied when $\mu\gg 1$ (adiabatic
limit) or $(\mu^3/\epsilon)^{1/4}\gg 1$ (sudden limit).

We assume that between the crossings only the phase evolves. The
propagator relative to this evolution reads
\begin{equation}\label{Uphase2}
   U_{ph}^{(2)}=\left(
         \begin{array}{cc}
           e^{-i\sigma/2} & 0 \\
           0 & e^{i\sigma/2} \\
         \end{array}
       \right),
\end{equation}
where $\sigma$ characterizes the adiabatic dynamical phase. As it is
found out in~\cite{Suominen92} and confirmed by our numerical
calculations, the propagator obtained in this way works very well
when ICA is satisfied. The total propagator in the interaction
picture can be written as
\begin{equation}\label{Upara2}
U_{Para}^{(2)}(+\infty,-\infty)=U_{LZ}^{T(2)}U_{ph}^{(2)}U_{LZ}^{(2)}=\left(
         \begin{array}{cc}
        \alpha & \beta \\
        -\beta^* & \alpha^*  \\
         \end{array}
       \right),
\end{equation}
where $U_{LZ}^{T(2)}$ indicates the transpose of~\eqref{LZProp}. The
reason of using the transpose lies in the fact that after the first
crossing one of the adiabatic states changes sign. Thus the
propagator at the second crossing needs to be written in the new
adiabatic basis. We need to change the signs of the non-diagonal
elements of~\eqref{LZProp}, or equivalently in our case, to take its
transpose.

In Eq.~\eqref{Upara2} we have
\begin{eqnarray}
   \label{alpha}
   \alpha &=& e^{i\sigma/2} [R^2+e^{-i(\sigma+2\phi)}(1-R^2)],\\
   \label{beta} \beta &=& 2iR\sqrt{1-R^2}\sin(\phi+\sigma/2).
\end{eqnarray}
Here $R$ and $\phi$ are obtained from Eqs.~(\ref{R}) and (\ref{LZphase}) by defining
\begin{equation}
   \Lambda=\frac{1}{2\sqrt{\epsilon\mu}},\qquad
   R=\exp\left[-\frac{\pi}{4\sqrt{\epsilon\mu}}\right].
\end{equation}
Additionally, the dynamical phase is obtained as
\begin{eqnarray}
   \sigma &=& 4\int_0^{\sqrt{\mu/\epsilon}} d\tau\,  \sqrt{(\epsilon \tau^2-\mu)^2+1}
          \nonumber\\
          &=& \frac{2\tau_c}{3}\left\{1+2\sqrt{\frac{\mu+i}{\mu}}\left(\mu\, \textrm{EE}
          \left[\arcsin\sqrt{\frac{\mu}{\mu-i}},\frac{\mu-i}{\mu+i}\right]\right.\right.
          \nonumber\\
          &&\left.\left.-i\textrm{EF}\left[\arcsin\sqrt{\frac{\mu}{\mu-i}},
          \frac{\mu-i}{\mu+i}\right]\right) \right\}.
          \label{sigma}
\end{eqnarray}
This is the full phase difference accumulated by the adiabatic
states between the two crossings. Here $\textrm{EE}$ stands for
$\textrm{EllipticE}$ and  $\textrm{EF}$ for $\textrm{EllipticF}$.
These are defined as $\textrm{EllipticE}[\alpha,m]=\int_0^\alpha
d\theta\,\sqrt{1-m\sin^2\theta}$ and
$\textrm{EllipticF}[\alpha,m]=\int_0^\alpha
d\theta\,1/\sqrt{1-m\sin^2\theta}$. Equation~(\ref{sigma}) shows
that the dynamical phase $\sigma$ can be written as a product of
$\tau_c$ and a function which only depends on $\mu$. When $\mu$ is
large we can approximate $\sigma\approx
8\mu^{3/2}/(3\sqrt{\epsilon})=4\tau_c\mu/3$, which correspond to the
value of $\sigma$ obtained in the absence of coupling. Therefore,
when the energy separation between the two levels at $t=0$ is large
compared with the coupling, it is possible to calculate the
dynamical phase in the diabatic basis. This approximation has been
used in Ref. \cite{Shimshoni91} in the derivation of an approximate
propagator for the parabolic model. However, when the exact result
(\ref{sigma}) is used, the resulting propagator works well also when
$v$ is not small compared with the energy separation at $t=0$.

As we noted with the LZ model, working in the interaction picture is
a necessity due to the otherwise infinite dynamical phases. If we
start either the LZ model or the parabolic model in a superposition
of states in the Schr\"odinger picture, problems will arise. Since
we are modelling here the idea of an interferometer, where the first
beamsplitter creates the superposition, we can avoid the infinite
phase problem by restricting our study to situations where only a
single eigenstate on the system is populated initially.

In a two-level model, starting from level $1$, the transition probability to level $2$ is now
\begin{equation}
   \label{P21} P^{(2)}_{1\rightarrow 2}=|\beta|^2=4R^2(1-R^2)\sin^2(\sigma/2+\phi).
\end{equation}
We note that the expression for $P^{(2)}_{1\rightarrow 2}$ given in
Ref.~\cite{Suominen92} contains misprints, the correct form is given
by Eq.~\eqref{P21}.

As an example we discuss the generalization of the parabolic model to three-level systems. The Hamiltonian is now
\begin{equation}\label{ParaHScaled3}
H_{Para}^{(3)}=\left(
         \begin{array}{ccc}
           2(a t^2-b) & \sqrt{2} v & 0 \\
           \sqrt{2} v&  0 & \sqrt{2} v\\
           0 & \sqrt{2} v &  -2(a t^2-b)
         \end{array}
       \right),
\end{equation}
Using Eq.~\eqref{Ugeneral} together with Eq.~\eqref{Upara2} we find the solution for a three-level system propagator as
\begin{equation}\label{DDD}
   U^{(3)}_{Para}=\left(
          \begin{array}{ccc}
            \alpha^2 & \sqrt{2}\alpha\beta & \beta^2 \\
            -\sqrt{2}\alpha \beta^{*} & |\alpha|^2-|\beta|^2 & \sqrt{2}\alpha^{*}\beta \\
            \beta^{*2} & -\sqrt{2}\alpha^{*}\beta^{*} & \alpha^{*2} \\
          \end{array}
        \right),
\end{equation}
where $\alpha$ and $\beta$ are given by Eqs.~(\ref{alpha}) and
(\ref{beta}). Starting again with level $1$ initially populated we
find that the transition probability to level $3$ is
\begin{equation}
   \label{P31} P^{(3)}_{1\rightarrow 3}=|\beta|^4=16R^4(1-R^2)^2\sin^4(\sigma/2+\phi).
\end{equation}
If we ignore the rather small $\Lambda$-dependence of $\phi$, we see
that the parameter $\Lambda$ controls the amplitude, and the
oscillations arise only from $\sigma$.

\section{Interferometry using spinor condensates}\label{Sec:Interf}

\subsection{Theoretical analysis}

The order-parameter of a spin-$F$ condensate can be written as
\begin{equation}\label{AAA}
   \psi(t)=\sum_{m_F=-F}^{F}\psi_{m_F}(\mathbf{r},t)\ket{F,m},
\end{equation}
with the normalization
$\sum_{m_F=-F}^F|\psi_{m_F}(\mathbf{r},t)|^2=n(\mathbf{r},t)$, where
$n(\mathbf{r},t)$ is the total particle density. We assume that the
condensate is confined in an optical dipole trap so that all the
components of the hyperfine spin can be trapped simultaneously, and
they are degenerate in the absence of any magnetic fields. In the
following we concentrate on an $F=1$ condensate. In the degenerate
case the time-evolution of an $F=1$ condensate is given by the
following set of time-dependent Gross-Pitaevskii
equations~\cite{Machida,Ho,Isoshima}
\begin{eqnarray}
     i\hbar\frac{\partial\psi_{1}}{\partial t}&=&{\cal L}\psi_{1}+
     \lambda_a\left(\psi_0^2\psi_{-1}^*+|\psi_1|^2\psi_{1}+
     |\psi_0|^2\psi_{1}\right.\nonumber\\
     &&\left.-|\psi_{-1}|^2\psi_{1}\right), \nonumber\\
           i\hbar\frac{\partial\psi_{0}}{\partial t}&=&{\cal L}\psi_{0}+
     \lambda_a\left(2\psi_1\psi_{-1}\psi_0^*+|\psi_{-1}|^2\psi_{0}+
     |\psi_1|^2\psi_{0}\right),\nonumber\\
     i\hbar\frac{\partial\psi_{-1}}{\partial t}&=&{\cal L}\psi_{-1}+
     \lambda_a\left(\psi_0^2\psi_1^*+|\psi_{-1}|^2\psi_{-1}+
     |\psi_0|^2\psi_{-1}\right.\nonumber\\
     &&\left.-|\psi_1|^2\psi_{-1}\right),
     \label{GPs}
\end{eqnarray}
with ${\cal L}=-\hbar^2 \nabla^2/(2m)+U(\mathbf{r},t)+\lambda_s
n(\mathbf{r},t)$, $\lambda_s= 4\pi\hbar^2 (a_0+2a_2)/(3m)$ and
$\lambda_a= 4\pi\hbar^2(a_2-a_0)/(3m)$. The $a_k$'s are the $s$-wave
scattering lengths in the scattering channels with total spin $k$.
The trapping potential is denoted by $U$, and it affects all spinor
components equivalently. The first term inside each parentheses
describes spin mixing processes where atoms in $m_F=+1$ and $m_F=-1$
states collide and create two atoms in $m_F=0$ state, or vice versa.
The number of particles in the $i$th component is  by definition
$N_i\equiv\int d^3r\, |\psi_i|^2$. From Eqs.~\eqref{GPs} we get
\begin{equation}
   \frac{\partial N_1}{\partial t}=\frac{\partial N_{-1}}{\partial
   t}=-\frac{1}{2}\frac{\partial N_0}{\partial
   t}=\frac{2\lambda_a}{\hbar} \int d^3 r\,
   \textrm{Im}[\psi_0^2\psi^*_{-1}\psi^*_1].
\end{equation}
It follows that
\begin{equation}
   \left|\frac{\partial N_0}{\partial t}\right|
   <\gamma N_0,\quad \gamma\equiv\frac{4|\lambda_a|n_{max}}{\hbar},
\end{equation}
where $n_{max}$ is the maximum density. Using the scattering lengths
relevant for $^{87}$Rb \cite{vanKempen02} and a maximum density
$n_{max}=10^{14}\textrm{ cm}^{-3}$ we find $\gamma\simeq 90\
s^{-1}$. If for example we have $\Delta t=100\ \mu$s the change in
the populations is less than 1 \%. The time-evolution given by the
Gross-Pitaevskii equations~\eqref{GPs} is then restricted to changes
in the phases of the spinor components. Thus we can describe it
defining the following effective propagator
\begin{equation}\label{GPprop}
U_{GP}^{(3)}=\left(
         \begin{array}{ccc}
           e^{i\theta_1} & 0 & 0 \\
           0 & 1 & 0 \\
           0 & 0 & e^{-i\theta_{-1}} \\
         \end{array}
       \right),
\end{equation}
where we set one of the three independent phases to zero. The phases
$\theta_1$ and $\theta_{-1}$ can be calculated by solving
Eqs.~\eqref{GPs}.

\subsection{The spinor interferometer}

We base our interferometer proposal for the analogy between optical
beamsplitters and the level crossings. The initial single-state
signal is split into two parts at the splitter-crossing. The two
parts travel along the different arms of the interferometer, and
combine again in the second splitter-crossing, to form a detectable
signal. Thus any difference in phase evolution along the two paths
is mapped into variation of populations. The closest counterpart of
our spinor interferometer is the quantum optical Mach-Zehnder
interferometer. In our case, however, the interferometer can have
more than two arms.

Let us consider a few points in our analogue. The splitting at the
level crossings can be controlled by the speed of change for the
bias field and by the coupling field strength. Thus we are not
limited to the 50-50 splitting. Here the controlling parameter is
the Landau-Zener term $R$. As for the phase evolution,
Eq.~\eqref{P31} demonstrates that it decouples from the splitting,
although we must remember that the standard phase evolution $\sigma$
will depend on the same parameters as $R$. On the other hand, we
have altogether three parameters [$a$, $b$ and $v$ in
Eq.~\eqref{ParaH}], so it is possible to fix $R$ and $\sigma$
independently.

The time-evolution of spinor states between the crossings represents
the internal part of the spinor interferometer. This is the time
region during which the phenomena to be investigated take place. As
discussed earlier, we consider temporal crossing separations which
are much shorter than the typical time scales of spin mixing
processes. Thus we suggest that the contribution~\eqref{GPprop} is
included to the total propagator, which becomes now
\begin{equation}\label{Finalresult}
   U_{TOT}^{(3)}=U_{LZ}^{T(3)}U_{ph}^{(3)}U_{GP}^{(3)}U_{LZ}^{(3)},
\end{equation}
where $U_{ph}^{(3)}$ is obtained from~\eqref{Uphase2} through
Eq.~\eqref{Ugeneral}. Thus the solution is obtained as a sequence of
separate events.

It is important to notice that the propagator~\eqref{Upara2} is
written in the adiabatic basis while~\eqref{GPprop} is written in
the diabatic basis [following Eqs.~\eqref{GPs}]. But the result
\eqref{Finalresult} is still well-defined. The reason lies in the
ICA approximation. Since $v\ll b$, the adiabatic and diabatic state
evolution coincide approximately outside the crossing regions. The
only problem arises in the vicinity of the crossings, but these time
intervals are much shorter ($t_Z\ll t_c$) and do not contribute much
to the phase evolution.

As an example let us start with the state $\ket{1,1}$ initially populated. The final
population of the state $\ket{1,-1}$ is
\begin{equation}\label{P1m1}
   P^{(3)}_{1\rightarrow -1}=16 R^4(1-R^2)^2\left\{\sin^4\chi
   +\cos2\chi\sin^2\Psi\right\},
\end{equation}
where
\begin{equation}
   \chi\equiv
   \frac{\sigma}{2}+\phi-\frac{\theta_1+\theta_{-1}}{4},
   \qquad\Psi\equiv\frac{\theta_1-\theta_{-1}}{4}.
\end{equation}
The result~\eqref{P1m1} depends on the parameters $\sigma$ and
$\phi$, which one can control externally, and on the GP phases
$\theta_1$ and $\theta_{-1}$. Thus the measure of the final
population of the  $\ket{1,-1}$ state (the output of the
interferometer) gives information on the Gross-Pitajevskii dynamics.
We can also see that the visibility of the interferometric
oscillations is maximized when $R=1/\sqrt{2}$.

A simple interferometric measurement would be thus to set
$R=1/\sqrt{2}$ to maximize the signal, then to vary $\chi$ (keeping
$R$ constant). As one can see from Eq.~\eqref{P1m1}, the difference
between minima and maxima gives information about $\Psi$.
Interestingly, possible fluctuations in the magnetic fields affect
the Zeeman shifts between adjacent spinor states equally, thus
affecting only $\sigma$.

One should note that because of the $\sin^4$-function, the fringes
are "sharper" than in the standard two-arm interferometer. If we go
to higher values of spin, such as $F=2$, the fringes become even
sharper. In analogy to diffraction patterns, the double slit is
replaced by a $2F+1$-slit. Note that in this case the number of
independent relative phases is $2F$.

\subsection{Experimental realization}

In this section we estimate the values of the parameter of the
interferometer looking at a real situation. The effect of the
interaction with an external magnetic field is described by the
general linear Zeeman model
\begin{equation}
   H_B^{(3)}= -g_F\mu_B\mathbf{B}\cdot\mathbf{\hat{F}},
\end{equation}
where $\mu_B$ is the Bohr magneton, $g_F$ is the Lande $g$-factor
(equal to 1/2 for $^{87}$Rb and $^{23}$Na),
$\mathbf{B}=(B_x,B_y,B_z)$ is the magnetic field, and
$\mathbf{\hat{F}}=(F_x,F_y,F_z)$ is the spin operator of a spin-1
particle (with $\hbar=1$). We can identify $H_B^{(3)}$ with
$H_{Para}^{(3)}$ defining $B_x=-4v/\mu_B$, $B_y=0$, and
$B_z=-4(at^2-b)/\mu_B$. With these definitions we get ($g_F=1/2$)
\begin{eqnarray}
   \nonumber
   \mu &=& \frac{|B_z(t=0)|}{B_x},\\
   \epsilon\mu &=& \left(\frac{2\hbar \dot{B}_z^c}{\mu_B B_x^2}\right)^2,\\
   \nonumber
    t_c &=& \frac{4|B_z(t=0)|}{|\dot{B}_z^c|},
\end{eqnarray}
where $|\dot{B}_z^c|$ is the rate of change of $B_z(t)$ at the crossings. The Zener time becomes
\begin{equation}
    t_Z =\left\{
    \begin{array}{ll}
    \frac{B_x}{2\dot{B}_c}, &\frac{4\hbar \dot{B}_z^c}{\mu_B B_x^2}\ll 1,\\
     & \\
    \sqrt{\frac{2\hbar}{\mu_B\dot{B}_z}}, &\frac{4\hbar \dot{B}_z^c}{\mu_B B_x^2}\gg 1.
    \end{array}\right.
\end{equation}
We estimate the values of $\epsilon$ and $\mu$ by choosing $B_x=60$
mG and $\dot{B}_z^c=5\cdot 10^4$ G/s, and $B_z(0)=300$ mG. These
give $\epsilon\approx 2$ and $\mu=5$. The distance between the
crossings is $t_c=24\ \mu$s while the Zener time is $t_{LZ}\approx
2\ \mu$s, which shows that the ICA is well satisfied, and our
approach should be applicable. From the previous values we also get
$R\approx 0.78$ which is very close to the value $1/\sqrt{2}$ which
maximizes the interferometric signal in~\eqref{P1m1}.

In our experimental scenario we have considered a constant coupling
field. Alternatively, we could use an rf-field scenario in analogy
to evaporative cooling~\cite{Pethick2008}. By performing the
rotating wave approximation and moving to a rotaing coordinate frame
we get a situation that is euivalent to the constant coupling field
scenario. Then the crossings take place when the bias-field-induced
Zeeman shifts become resonant with the rf-frequency, and the bias
field does not have to pass a zero-point value. Yet another
variation would involve chirping of the rf-frequency, as it can also
be described as time-dependent variation of the energy of the spinor
states.

\section{Conclusions}

In this paper we have presented the basic idea of a spinor
interferometer. It is based on applying time-dependent magnetic
fields to provide level crossings of spinor states in analogy to
optical beam splitters. Interferometer-like studies that apply the
Landau-Zener model twice and show path-related interference are not
uncommon in physics, see e.g. Ref.~\cite{Sillanpaa06}. The spinor
interferometer is a partial extension into the case of a multi-arm
interferometer, which is perhaps harder to realize in quantum
optics. The extension is only partial due to the nature of the
spinor. For spin-$F$ we get $2F+1$ paths and $2F$ independent
phases, but the spinor dynamics is nevertheless characterized by the
same small number of parameters as the corresponding two-state
model, as the Majorana representation shows. Instead of being a
limitation, one can see this spinor property as an advantage when
considering possible error sources.

It should also be clear that one is not limited to the parabolic
model. We have chosen to use it because then we can link the
intuitive concept of using the LZ model twice within ICA to the full
solution of the model in the adiabatic limit, as done in
Ref.~\cite{Suominen92}. Our next step is to consider more carefully
how the dynamics of Eqs.~\eqref{GPs} appear in the phase evolution
given by Eq.~\eqref{GPprop}.

\begin{acknowledgments}
The authors acknowledge the Fondazione A. Della Riccia and Finnish
CIMO (R.V.), EC projects CAMEL and EMALI (H.M), and the Academy of
Finland projects 108699 and 115682 (H.M and K.-A.S) for financial
support. The authors thank Boyan Torosov and Nikolay Vitanov for
enlightening discussions.

\end{acknowledgments}

\end{document}